# What's the point of documentation?

Louise Pryor

louise@louisepryor.com

ABSTRACT

*We give a brief characterisation of the purposes and forms of documentation in and of spreadsheets.*

## 1. INTRODUCTION

It is a truth universally acknowledged, that software documentation [Lethbridge et al, 2003] is a Good Thing, and spreadsheets are no exception [Morison & Jordan, 2000]. The FSA, in a recent newsletter [FSA, 2006], described what they had seen in the way of good practice for financial modelling systems: "Acceptable standards of documentation were established, agreed by the firm, and themselves documented." They went on to say "The standards of control and documentation applicable to systems developments applied equally to spreadsheets".

But what is documentation *for*? It seems to be more or less assumed that *all* documentation is worth while, and the more documentation the better. However, given that most financial models and spreadsheets are developed with limited resources, and writing documentation takes time, it's important to consider what forms of documentation are most useful and productive. In order to do this we must think about what we are trying to achieve through the production of documentation.[1]

## 2. WHAT DOCUMENTATION DOES

Spreadsheet documentation may do any of the following:

- **Specify** the intended working of the spreadsheet
- **Record** what was done
- **Explain** how the spreadsheet works
- **Instruct** the user how to use or update the spreadsheet

Ideally, a spreadsheet would have documentation serving all four purposes.

A specification may be anything from a single sentence to a separate, long, document. In essence, it describes the theory of the spreadsheet rather than the implementation.

A record of what was done does not necessarily contribute much to the understanding of the current state of the spreadsheet, if it describes changes that were made much earlier in its life.

An explanation describes the implementation of the spreadsheet.

Instructions are often used to remind users how and when to perform manual processes, such as running a macro, or what inputs are required.

---

[1] Although the primary focus of this paper is on documentation of spreadsheets, most of the points also apply to other financial models and indeed to user developed applications in general.

Documentation is usually easier to understand if it is written for a single purpose. It really helps a user if instructions are written (and labelled) as instructions, rather than buried deep in the record of a change.

## 3. WHO IS DOCUMENTATION FOR?

Like the spreadsheet itself, documentation may have the following types of user:

- A **Viewer** looks at the results, but makes no changes. A Viewer may never see the actual spreadsheet, but only printouts of selected parts, but they are still using the spreadsheet.

- A **Player** changes input cell values, but doesn't change formulae and layouts. They may execute macros.

- A **Changer** changes formulae and makes minor layout and formatting changes. They correct and enhance the spreadsheet, without making major changes.

- A **Developer** makes major changes, writes VBA code, designs and implements the spreadsheet from scratch.

- An **Auditor** may change input values. They must understand the working of the spreadsheet and be able to trace the inputs and outputs.

A **Reviewer** may share the needs of any of the other types of users, depending on what aspects of the spreadsheet they are reviewing. A **Tester** [Pryor, 2004] is essentially like a Player: their role is to check the operation and results of the spreadsheet, without making changes to its operation. For simple spreadsheets there may be no effective difference between many of these roles.

The same person may use the spreadsheet in several different ways during its lifetime. Users are characterized by the role they are playing at the time, rather than by what role they may be capable of playing.

## 4. WHAT DOCUMENTATION HELPS WHO?

There is no type of documentation that is *always* useless to any type of user, but in general some types are more useful than others. The following table summarizes the general utility of the different types of documentation to different users:

|           | Specify | Record | Explain | Instruct |
|-----------|---------|--------|---------|----------|
| Viewer    | X       | x      | X       | X        |
| Player    | X       | x      | X       | X        |
| Tester    | X       | x      | x       | X        |
| Changer   | X       | X      | X       | X        |
| Developer | X       | X      | X       | X        |
| Auditor   | X       | X      | X       | X        |

The table can form the basis of a useful checklist when reviewing the presence and adequacy of documentation.

## *Specify*

A fairly abstract level of specification is useful to everybody: "This spreadsheet models the effect of inflation on sales" for example. Some users need rather more detail than that: a Changer or Developer needs full details of the theoretical model that should be used, so that they can implement it, and a Tester needs those details so that they can see whether it has been implemented. You can't tell whether a spreadsheet is doing the right thing unless you know what the right thing is. Even a Viewer or Player may need to know the theory behind the spreadsheet so that they can interpret the results.

## *Record*

A record of what was done is especially useful to an Auditor, but is also helpful to Changers and Developers, who may need to work out why things are going wrong. However, a record is rarely a good substitute for either explanation or instruction. Recording documentation may be a simple narrative of steps taken, or a more formal record of versions, changes made, reviewers, tests performed, and so on. Information about the sources of data or parameters may count as either recording or explaining documentation.

## *Explain*

Explanations of how the spreadsheet is put together, or why specific design decisions were made are usually primarily intended for Changers and Developers, but are also useful to Auditors. Explanations of the sources of data or parameters may be useful for all users. A simple narrative of "what I did when building this spreadsheet" is rarely useful as an explanation, especially as any information it gives may be superseded later on in the narrative. Explanations of the significance of outputs are useful to all users, especially Viewers or Changers who may not have the skills necessary to infer what is going on from the formulae or code.

## *Instruct*

Instructions may be directed at any type of user. They are especially important for Viewers and Players, who may not always be able to infer what should be done from the structure or code in the spreadsheet itself.

### 5.    FORMS OF DOCUMENTATION

Documentation may take many different forms:

- A **separate document** is often used for long, formal specifications that are themselves subject to review and sign-off. Records of changes and versions may also be kept in a separate document or database. The potential disadvantage of a separate document is that it may be difficult to maintain consistency between the documentation and the spreadsheet.

- **Implicit** documentation is widely used. The names of worksheets, ranges and cells, and modules and variables in VBA code come into this category. Formatting may also provide documentation, for example if colours are used consistently to indicate which cells are inputs, or to indicate potential errors.

- Documentation **within the code** is perhaps the most common form. It is particularly easy to use in spreadsheets (compared to other types of software) as it often simply takes the form of text in cells. It is well suited to instructions and some types of explanation, as it can be placed close to the cells to which it refers.

- Documentation as a **separate block** in code, for example as a separate worksheet, can be very useful, especially for keeping records.

- Documentation in the **user interface** overlaps with documentation within the code for spreadsheets, but is clearly distinct in more conventional software. It includes text in user forms, text boxes and other images.

The most appropriate documentation method depends on the type of documentation and the user for which it is intended, as well as the culture in which it will be used. If your team commonly uses separate documents or centralized systems for specific types of documentation, then you should conform to the common practice.

## 6.  WHO BENEFITS FROM DOCUMENTATION?

Just about everybody benefits from clear, accurate documentation. The benefits of out of date documentation written without a specific purpose in mind are less obvious, and indeed are often non-existent. To get the most out of documentation, the following should be true:

- It should be written specifically as specification, record, explanation or instruction. That way it will be easy for the reader to understand.

- It should be easily available to anybody who wants it; this often (but by no means always) means that it should be part of the spreadsheet that it applies to.

- It should be kept up to date, otherwise it might mislead the reader.

- When writing documentation, it is often helpful to bear in mind the specific type of user for whom you are writing.

It is not only the reader who benefits from documentation. The writer often gains a lot, too; if the writer is a developer, writing some of the documentation in advance can help to focus the mind and prevent false starts. Articulating your ideas can save you from many dead ends. The documentation process can often throw up bugs in the spreadsheet or ambiguities in the specification, especially if the writer is a user other than a developer of the spreadsheet.

Appropriate documentation will help people other than the developer have confidence in the results of a spreadsheet. They will be to able to tell what the spreadsheet is intended to do, how it does it, what data it uses, how to use it and interpret the results, and what tests and reviews have been performed.